\documentclass{iopart}
\usepackage{cite}

\jl{6}        
\eqnobysec    

\def\beq{\begin{equation}}
\def\eeq{\end{equation}}

\def\dualp#1{{}^{\ast_{(\hbox{$\scriptstyle #1$})}} \kern-1pt}
\def\dual{\,{}^\ast{}\kern-1.5pt}

  \def\del{\nabla}
\def\div{\mathop{\rm div}\nolimits}    \def\curl{\mathop{\rm curl}\nolimits}
\def\Scurl{\mathop{\rm Scurl}\nolimits} 
\def\TF{{}^{(\rm TF)}}
\def\SYM{\mathop{\rm SYM}\nolimits}  

\def\rmd{{\rm d}}
\begin{document}

\title[The Simon and Simon-Mars Tensors for Stationary Einstein-Maxwell Fields]
{The Simon and Simon-Mars Tensors for Stationary Einstein-Maxwell Fields}

\author{
Donato Bini${}^{\dag\,\ddag}$,
Christian Cherubini${}^{\ast,||}$,
Robert T. Jantzen${}^{\S \,\ddag}$,
Giovanni Miniutti${}^{\P \,\ddag}$
}
\address{
  ${}^{\dag}$\
Istituto per le Applicazioni del Calcolo \lq\lq M. Picone\rq\rq, C.N.R.,
   I-- 00161 Roma, Italy
}
\address{
  ${}^{\ddag}$\
  International Center for Relativistic Astrophysics,
  University of Rome, I--00185 Roma, Italy
}
\address{
${}^{\ast}$
  Department of Physics \lq\lq E.R. Caianiello\rq\rq, University of Salerno, I--84081, Italy
}
\address{
${}^{||}$\
Institute of Cosmology and Gravitation, University of Portsmouth,
Portsmouth, PO1 2EG, UK}
\address{
  ${}^{\S}$\
  Department of Mathematical Sciences, Villanova University,
  Villanova, PA 19085, USA
}
\address{
  ${}^{\P}$\
Institute of Astronomy, University of Cambridge, Madingley Road, Cambridge, CB3 OHA 
}

\begin{abstract}
Modulo conventional scale factors, the Simon and Simon-Mars tensors are defined for stationary vacuum spacetimes so that their equality follows from the Bianchi identities of the second kind. In the nonvacuum case one can absorb additional source terms into a redefinition of the Simon tensor so that this equality is maintained. Among the electrovacuum class of solutions of the Einstein-Maxwell equations, the expression for the Simon tensor in the Kerr-Newman-Taub-NUT spacetime in terms of the Ernst potential is formally the same as in the vacuum case (modulo a scale factor), and its vanishing guarantees the simultaneous alignment of the principal null directions of the Weyl tensor, the Papapetrou field associated with the timelike Killing vector field, the electromagnetic field of the spacetime and even the Killing-Yano tensor.
\end{abstract}

\pacno{04.20.Cv}

\section{Introduction}

In the $1+3$ gravitoelectromagnetic approach to general relativity associated with splitting spacetime quantities using a unit timelike vector field $u$, the 4-velocity of an observer congruence,
the Bianchi identities for the curvature tensor take a Maxwell-like form for the electric and magnetic parts of the Weyl tensor with sources arising from the Ricci tensor terms, which in turn can be expressed in terms of the energy-momentum tensor of the spacetime through the Einstein field equations \cite{Ellis73,Maartens,Ellis99}.
For the stationary case with $u$ directed along the timelike Killing vector field, these equations are greatly simplified and suggested the introduction of the complex Simon tensor \cite{Simon}, which in the vacuum case equals the simpler Simon-Mars tensor \cite{Mars99,Mars00,Mars01,Fasop99, Fasop00,cys1} (modulo conventional scale factors). 
The Simon-Mars tensor is a natural algebraic combination of the self-dual fields associated with the Papapetrou field of the timelike Killing vector field and the Weyl tensor. 

Simon introduced his Simon tensor for vacuum stationary spacetimes as a complex generalization of 
the Cotton-York tensor, to which it reduces in the static case, where its vanishing guarantees that the space sections are conformally flat, as occurs in the Schwarzschild spacetime.
Its vanishing also characterizes 
the Kerr spacetime within certain classes of solutions of the stationary vacuum Einstein equations. 
Mars has given an equivalent form of the Simon tensor (the Simon-Mars tensor) that has been very useful in understanding the meaning of its vanishing in the Kerr spacetime, where it leads to the alignment of the principal null directions of the Weyl tensor of the spacetime with those of the Papapetrou field associated with the timelike Killing congruence.

Here we extend the gravitoelectromagnetic analysis of the vacuum case \cite{cys1}
to electrovac stationary spacetimes which contain a source-free electromagnetic field, 
where the Simon tensor and the Simon-Mars tensor defined by the same formulas as in vacuum  differ by additional source terms. However,
these source terms can be absorbed into a redefinition of the Simon tensor so that it continues to be equivalent to the Simon-Mars tensor defined exactly as in the vacuum case, and of course having all the same tensor symmetries. 
Remarkably for the Kerr-Newman spacetime and even the more general family of Kerr-Newman-Taub-NUT spacetimes,  the  situation is very similar to the Kerr spacetime. First the vanishing of the Simon-Mars tensor (and hence of the newly defined Simon tensor) can still be explained in terms of the simultaneous alignment of the principal null directions of the Weyl tensor, the Papapetrou field associated with the timelike Killing vector field, the electromagnetic field of the spacetime and even the Killing-Yano tensor. Secondly when written in terms of the Ernst potential, the newly defined Simon tensor including the source terms formally coincides with the vacuum expression (modulo a scale factor).

Our extension of the Simon tensor to stationary electrovac spacetimes takes a different approach from the work of Marklund and Perj\'es on stationary perfect fluid spacetimes \cite{Perjes}. They are also able to incorporate additional source terms into the definition of the nonvacuum generalization of the Simon tensor in such a way that the symmetry properties of the tensor are respected. However, they focus on the spatial Bianchi identities (for the conformally rescaled spatial metric) rather than the spacetime Bianchi identities.

Finally it is worth noting that a prime motivation of our discussion, begun in \cite{cys1}, is to make more accessible the geometry surrounding the Simon tensor, whose existing literature assumes detailed knowledge of stationary exact solution techniques that outsiders simply do not possess. Nonexperts cannot even recognize the equivalence of different formulas for the Simon tensor stated by various experts in these manipulations. Appendix A reconciles these definitions, while the main body recasts the geometry in the natural complex version of the Maxwell-like equations for the Weyl tensor pioneered by Ellis \cite{Ellis73,Maartens,Ellis99}.

\section{The $1+3$ Maxwell-like equations for the Weyl tensor with sources}

The Bianchi identities of the second kind equate the divergence of the Weyl curvature tensor $C_{\alpha\beta\gamma\delta}$ to half the Cotton tensor \cite{Cotton}, which can be thought of as a current in analogy with electromagnetism \cite{HE}
(where $F^{\alpha\beta}{}_{;\beta} = 4\pi J^\alpha$)
\beq
\label{current}
 C^{\alpha\delta}{}_{\beta\gamma;\delta}
=  -\del_{[\beta} ( R^\alpha{}_{\gamma ]}  -\frac16 R \delta^\alpha{}_{\gamma ]} )
\equiv J^\alpha{}_{\beta\gamma}\ ,
\eeq
following the conventions of Misner, Thorne and Wheeler \cite{mtw}, including the signature 
$-$+++.
The Einstein equations (in Ricci form) 
\beq\label{eq:EEricci}
  R^\alpha{}_\beta 
 = \kappa(T^\alpha{}_\beta - {\textstyle\frac12} T^\gamma{}_\gamma\delta^\alpha{}_\beta)
\eeq
can then be used to replace the Ricci tensor and scalar curvature terms in this current by the energy-momentum tensor.
Splitting these equations with respect to a generic
timelike congruence $u$ leads to their $1+3$ Maxwell-like form  \cite{Ellis99}
given explicitly in our notation in \cite{cys1}.

The Weyl tensor splits into two symmetric tracefree spatial fields,  its electric and magnetic parts respectively
\beq
\fl\qquad
  E \,^{\alpha} {}_{\beta} 
=  C^{\alpha} {}_{\gamma\beta\delta}\, u^\gamma \, u^\delta \ ,\quad
  H \,^{\alpha} {}_{\beta} 
= - {}^\ast C^{\alpha} {}_{\gamma\beta\delta}\, u^\gamma \, u^\delta
= \frac{1}{2}\, \eta \, ^{\alpha} {}_{\gamma} {}^{\delta}\,
 C^{\gamma}{}_{\delta\beta\rho}\, u^\rho 
\ .
\eeq
If $a$ is the acceleration vector and $\omega$ the vorticity vector of $u$,
introducing the complex spatial fields from the gravitoelectromagnetic connection and Weyl curvature fields
\beq
    z=-a-i\omega = g-i\vec H/2 \ ,\qquad
    Z=E-iH\ ,
\eeq
where $g$ and $\vec H$ are the gravitoelectric and gravitomagnetic vector fields,
leads to a more efficient representation of those equations.
These complex fields are associated with the $O(3,C)$ representation of the Lorentz group exploited in the `complex vector / self-dual formalism' \cite{Kramer,Debeve,Taub} in turn closely related to the Newman-Penrose formalism.
Here we specialize to the case of a stationary electrovacuum spacetime
and align $u = M^{-1}\xi$ with the associated Killing vector $\xi$, so that
the acceleration can be expressed as $a = \nabla \ln M$, where 
$M=|-\xi_\alpha \xi^\alpha|^{1/2}$. 

Splitting the Weyl divergence equations and expressing them in terms of an adapted frame $e_\top =u, e_a$, $a=1,2,3$, where the spatial frame $e_a$ is orthogonal to $u$, 
one finds in an index-free notation the identities \cite{cys1}
\beq\fl\qquad
\label{bianchi}
M^{3}\div [M^{-3} Z] = 3 z \cdot Z +  \rho^{(G)}
\ ,\quad
M^{-1}\Scurl [MZ] =  z\times Z + i J^{(G)}
\ ,\eeq
where the complex current fields are defined by
\beq
\fl\qquad
\rho^{(G)}_a=J^\top{}_{a\top}+iJ^{*} {}^\top{}_{a\top}\ , \qquad 
J^{(G)}_{ab}=-[J_{(ab)\top}+i J^{*}{}_{(ab)\top}] \ .
\eeq
The div and Scurl spatial operators on symmetric spatial 2-tensors $S$ 
\beq
  [\div(u) S]^\alpha = \del(u)_\beta S^{\alpha\beta}\ ,\
  [\Scurl(u) S]^{\alpha\beta} 
     = \eta(u)^{\gamma\delta(\alpha} \nabla(u)_\gamma S^{\beta)}{}_\delta
\eeq
are defined in terms of the
spatial covariant derivative $\nabla(u)$, the spatial volume 3-form $\eta(u)_{\alpha\beta\gamma} 
=u^\delta \eta_{\delta\alpha\beta\gamma}$ 
(associated with the spatial duality operation $\dualp{u}$, used for example to define the vorticity vector $\omega(u)^\alpha=\frac12\eta(u)^{\alpha\beta\gamma}\nabla_\beta u_\gamma$ from the corresponding 2-form, see appendix A of \cite{cys1} for sign conventions) and the spatial cross and dot products
associated with $u$, 
all given explicitly in \cite{cys1} by spatially projecting spacetime derivatives and quantities orthogonally to $u$. 
For example, the cross-product of a spatial vector and spatial symmetric tensor is
\beq
  [X \times_u A]^{\alpha\beta} = \eta (u)^{\,\gamma\delta (\alpha }
  X_{\gamma} A^{\beta)} {}_{\delta}\ .
\eeq
For simplicity we drop the reference to $u$ on these spatially projected operators, resorting to the abbreviation $\vec\nabla=\nabla(u)$ to distinguish the spatial covariant derivative from the spacetime covariant derivative $\nabla$,
an unnecessary distinction when acting on stationary scalars. In an adapted frame spatial quantities may be expressed using only Latin indexed components.

The spatial cross product of $z$ and $Z$ which appears in the Scurl equation in the projected Bianchi identity (\ref{bianchi}) defines the (complex, symmetric and tracefree) Simon-Mars 2-tensor
\beq
\label{simon,cys1}
{\rm SimonMars} = z\times Z\ .
\eeq
By using that Scurl equation to define the nonvacuum Simon 2-tensor for any stationary spacetime (with any energy-momentum source) as
\beq
  {\rm Simon} = M^{-1}\Scurl [MZ] - i J^{(G)}\ ,
\eeq
this identity then guarantees that the Simon and Simon-Mars tensors also remain equal in the nonvacuum case. Of course once the Ricci tensor in the source term $J^{(G)}$ is replaced using the Einstein equations (\ref{eq:EEricci}), this equality between Simon and Simon-Mars is only true if they are each evaluated on a solution of those equations.
Apart from a scale factor, our Simon tensor is the spatial dual of the three-index spacetime tensor introduced by Simon in the vacuum case \cite{Simon}, as shown in Appendix A.

In the special case in which the source term has the form $J^{(G)} = -i \nabla \ln\sigma \times Z$, then the Scurl identity for any scalar $\sigma$ and any symmetric spatial 2-tensor $A$
\beq
\sigma\Scurl [\sigma^{-1} A]=-\vec\nabla \ln\sigma \times A + \Scurl A 
\eeq
enables it to be incorporated into the existing Scurl term
\beq\label{eq:simonsigma}
{\rm Simon} = (\sigma/M)\Scurl [\sigma^{-1} M Z]  \ .
\eeq
This again would re-establish the vacuum property that the Simon tensor is proportional to a Scurl, but with a generalized Simon tensor incorporating the source terms so that it still equals the Simon-Mars tensor. This is the case with the family of Kerr-Newman-Taub-NUT electrovac spacetimes (including the more physically interesting Kerr-Newman spacetime), where the extra factor of $\sigma$ converts this to the same Scurl expression when written in terms of the Ernst potential as in the vacuum formula (modulo a scale factor).

The vanishing of the Simon-Mars tensor ($z\times Z=0$) is equivalent to
the alignment of $z$ and $Z$ in the sense that $Z$ must be proportional to the tracefree tensor product of $z$ with itself \cite{cys1}
\beq
Z=k[z\otimes z]^{\rm(TF)} \ ,
\eeq
which is an algebraic consequence of the definition of the cross product for symmetric spatial 2-tensors \cite{cys1}. Thus the vanishing of the generalized Simon tensor corresponds to the alignment of $z$ and $Z$ as in the vacuum case.

\section{Maxwell equations and the Papapetrou field}

The stationary Maxwell-like equations (\ref{bianchi}) for the Weyl tensor expressed in complex self-dual form 
are very similar to the stationary  Maxwell equations expressed in complex form \cite{cys1}
\beq
\fl\qquad
\label{eq:max}
M^{-2}\div(M^2 x) = -2 z\cdot x+4\pi \rho \ ,\qquad
M^{-1}\curl(M x)=4\pi i J
\eeq
satisfied by the spacetime's electromagnetic field (using an arrow to distinguish the electric field
from the electric part of the Weyl tensor in index-free notation) 
\beq
 F=u\wedge \vec E + \dualp{u} B\ ,\qquad x = \vec E -iB\ .
\eeq
They only differ in form by the missing $z\times x$ term in the curl equation here, where $\dualp{u}B$ is the spatial dual of the magnetic 1-form field $B$.
However, when $z$ and $Z$ are aligned (i.e., $z\times Z=0$) as in the Kerr-Newman-Taub-NUT  spacetimes, the correspondence is complete. The sources $\rho$ and $J$ for the spacetime electromagnetic field are zero for the electrovac spacetimes considered here.
The curl equation for the electromagnetic field implies that $Mx$ is the gradient of a potential
\beq\label{potx}
Mx=\nabla \Phi\ , \quad{\rm or}\quad  x=M^{-1}\nabla \Phi\ .
\eeq 

Papapetrou \cite{pap66} observed that a Killing vector field $\xi$ can be
interpreted as a Lorentz gauge vector potential for an electromagnetic field later called the Papapetrou field \cite{Fasop99,Fasop00}
\beq\label{eq:papfield}
\widetilde F_{\alpha\beta}
=[d \xi]_{\alpha\beta}
= \nabla_\alpha \xi_\beta - \nabla_\beta \xi_\alpha
= 2\nabla_\alpha \xi_\beta 
\eeq
satisfying Maxwell's equations with an effective current source proportional to the Ricci tensor
\beq\label{eq:papst}
\fl\qquad
 \xi_{(\alpha;\beta)}=0 
\rightarrow
 \xi^\alpha{}_{;\alpha}=0 
\rightarrow
 \xi_{\alpha;\beta}{}^{;\beta} = -\xi_\beta R^\beta{}_\alpha 
\rightarrow
\tilde F_{\alpha\beta}{}^{;\beta} = 2\xi_\beta R^\beta{}_\alpha \ ,
\eeq
where this divergence condition is merely the trace of the Ricci identity for a Killing vector field.
Rescaling the Papapetrou field $\tilde F$ by the factor $(2M)^{-1}$ \cite{cys1}
and introducing its complex self-dual representation yields the 
self-dual 2-form $\mathcal{F}$ associated with $z$
 \beq\label{eq:Papselfdual}
\mathcal{F} = (2M)^{-1}(\tilde F + i \dual \tilde F) 
  = u\wedge z +i\dualp{u} z \ ,
\eeq
which when multiplied by $M$
satisfies the same Maxwell equations but with nonzero sources generated by the electromagnetic field. These equations
\beq
\label{eqpap}
\fl\qquad
 M^{-3}\div(M^3z)=-2 z\cdot z- x\cdot \bar x\ ,\ \quad
 M^{-2} \curl(M^2 z) =  x \times \bar x \ .
\eeq
are obtained by splitting the vector Eqs.~(\ref{eq:papst}) and
using the Einstein equations (\ref{eq:EEricci}) to replace the Ricci tensor appearing there in terms of the appropriate projections of the electromagnetic energy-momentum tensor of the spacetime
\beq
\kappa T^{\top}{}_{\top} = - x\cdot \bar x\ ,\quad
 i \kappa T^{\top}{}_{a} =  [x \times \bar x]_a \ ,
\eeq
which then leads to the source terms appearing in these Maxwell equations.

Inserting the potential representation (\ref{potx}) of $Mx$ into the rescaled Papapetrou field curl equation of Eqs.~(\ref{eqpap}) then yields
\beq
M^{-2} \curl(M^2 z) = M^{-2} \nabla \Phi \times \nabla \bar\Phi =- M^{-2} \curl (\bar \Phi \nabla \Phi ) \ ,
\eeq
namely 
\beq\label{eq:MPhi}
\curl(M^2 z + \bar \Phi \nabla \Phi)=0\ .
\eeq
This latter quantity in parentheses must therefore be a gradient, defining
the Ernst potential $\mathcal{E}$
\cite{Kramer} (see their Eq.~(18.34) with $F=-M^2$, reducing to the integrated Eq.~(18.29) in the vacuum case $\Phi=0$) modulo a numerical factor
\beq\label{eq:ernstpot}
\nabla \mathcal{E} = -2 (M^2 z + \bar \Phi \nabla \Phi) \ .
\eeq

The remaining divergence equation of Eqs.~(\ref{eqpap}) is the Ernst equation, which can be expressed in the form (equivalent to Eq.~(18.38) of \cite{Kramer}, see Appendix A)
\beq
\label{ernst}
M \div (M\nabla \mathcal{E} ) = - 2M^2 z\cdot \nabla \mathcal{E}\ .
\eeq
This result follows from 
using the definition of $\mathcal{E}$ to  re-express $x$
in terms of the gradient of $\Phi$ in the two divergence equations (\ref{eq:max}) and (\ref{eqpap}) 
\beq
\fl\qquad
\frac12 \div \nabla \mathcal{E} +\bar \Phi \div \nabla \Phi = M^2 (a+2z)\cdot z, \qquad
\div \nabla \Phi = - (a+2z)\cdot \nabla \Phi
\eeq
and using the second of these equations to eliminate $\div \nabla \Phi$ in the first, yielding
\beq
\div \nabla \mathcal{E} +a \cdot \nabla \mathcal{E}= - 2z\cdot \nabla \mathcal{E}\ ,
\eeq
which can then be rewritten as Eq.~(\ref{ernst}).

\section{Simon and the Ernst potential}

The Simon-Mars tensor can be evaluated in a form which shows its relation to the Ernst potential expression for the Simon tensor, assuming that the Einstein field equations are satisfied so that the two tensors are in fact equal.
Following the notation of \cite{cys1},
the purely spatial components and the mixed time-space components of the Ricci tensor field equations (see \cite{Kramer}, Eqs.~18.15 and 18.16) are respectively
\begin{eqnarray}
\label{ricci}
[{}^{(4)} R_{ab}]^{\rm (TF)}
&=& -2M^{-2}[\nabla_{(a}\Phi \nabla_{b)}\bar\Phi]^{\rm (TF)}\ ,\nonumber \\
{}^{(4)}R_{\top b} 
&=& iM^{-2}[\curl (M^2 z)]_b 
   =-iM^{-2}[\curl (\bar \Phi \nabla \Phi)]_b \ ,
\end{eqnarray}
where $z$ and $\Phi$ are related to the Ernst potential by (\ref{eq:ernstpot}).
A straightforward computation gives the electric and magnetic part of the Weyl tensor in terms of the gravitoelectromagnetic fields $a$ and $\omega$ and the Ricci tensor of the quotient 3-manifold \cite{cys1}
\begin{eqnarray}
E&=& -\frac12 [4 \omega \otimes \omega -\vec\nabla a -a\otimes a -^{(3)}{\rm Ricci}]^{\rm (TF)}\ , \nonumber \\
H&=&-\SYM [\vec\nabla \omega +2 a \otimes \omega]^{\rm (TF)}\ .
\end{eqnarray}
Then using the formula
\beq
\label{R3vsR4}
[{}^{(3)}{\rm Ricci}]^{\rm (TF)}
=[{}^{(4)}{\rm Ricci} + \vec\nabla a + a\otimes a +2 \omega \otimes \omega]^{\rm (TF)}\ ,
\eeq
one finds
\beq
\label{Zdef}
Z = E-iH
  =[-\SYM(\vec\nabla z) + z \otimes z + \frac12 {}^{(4)}{\rm Ricci}]^{\rm (TF)}\ .
\eeq

The Simon-Mars tensor is then
\beq
{\rm SimonMars}
  = z\times (-\SYM \,\vec\nabla z+\frac12 {}^{(4)}{\rm Ricci})
\eeq
since $z\times(z\otimes z)$ is identically zero. 
By next replacing the first term here using the Scurl identity valid for an arbitrary spatial vector field $X$
\beq
\Scurl  (X\otimes X) 
= -X\times \SYM  (\vec\nabla X) + \frac32 [\SYM (X\otimes \curl X)]^{\rm (TF)}\ ,
\eeq
and using the Einstein Eqs.~(\ref{ricci}) to replace the spatial components of the spacetime Ricci tensor, the Simon-Mars tensor becomes
\begin{eqnarray}
\label{nuovosimon}
{\rm SimonMars}
&=& \Scurl (z\otimes z) -\frac32 [\SYM (z\otimes \curl \,z)]^{\rm (TF)}
\nonumber\\ &&\quad
 -M^{-2} z\times  \SYM (\nabla \Phi \otimes \nabla\bar\Phi )\ .
\end{eqnarray}
Finally by substituting $z=M^{-2}(M^2 z)$ into the curl term here expanding the result with the appropriate product rule and incorporating the resulting additional acceleration term into the Scurl term, and then using Eq.~(\ref{eq:ernstpot}) 
\beq
\curl (M^2 z) =-\curl (\bar\Phi \nabla \Phi)
\eeq
to substitute the other term in this expansion, one finds
\begin{eqnarray}
\label{nuovosimon1}
{\rm SimonMars}
&=& M^{-3}\Scurl (M^3 \, z\otimes z)
 - \frac32 [\SYM (z\otimes (x\times \bar x)]\TF \nonumber\\
 && \quad 
-  z\times  \SYM (x \otimes \bar x ) \ .
\end{eqnarray}

In the vacuum case where $x=0$, this reduces to
\beq
\label{nuovosimon2}
{\rm SimonMars} = M^{-3}\Scurl (M^3 \, z\otimes z)\ , 
\eeq
which can be rewritten in terms of the Ernst potential by making the substitution 
$z= -\frac12 M^{-2} \nabla \mathcal{E}$
\beq
\label{simonmarsvacuum}
{\rm SimonMars}
= \frac14 M^{-3}
  \Scurl (M^{-1}\nabla \mathcal{E} \otimes \nabla \mathcal{E})
\ .
\eeq
As shown in appendix A, this is proportional to the form of the Simon tensor given in \cite{Kramer} and to the original form given by Simon. 

\section{The Kerr-Newman-Taub-NUT spacetime}

In  Boyer-Lindquist coordinates $(x^\alpha)=(t,r,\theta,\phi)$, with the abbreviations
$(c,s)=(\cos\theta,\sin\theta)$,
the (exterior) Kerr-Newman-Taub-NUT (KNTN)
spacetime the line element is
\cite{DemNew,Mil,NUT,McGRuf,DodTur}
\begin{eqnarray}
\label{metrica}
\rmd s^2
   &=&-\frac{1}{\Sigma}(\Delta -a^2 s^2)\rmd t^2
    +\frac{2}{\Sigma}[\Delta \chi -a(\Sigma +a \chi)s^2]\rmd t
\rmd \phi \nonumber \\ 
    &&+\frac{1}{\Sigma}[(\Sigma +a \chi)^2 s^2 -\chi^2 \Delta ]\rmd \phi^2
    +\frac{\Sigma}{\Delta}\rmd r^2 +\Sigma d\theta^2\ ,
\end{eqnarray}
and the corresponding electromagnetic Faraday tensor can be expressed in terms of the 2-form
\begin{eqnarray}
\label{max}
F&=&\frac{Q}{\Sigma^2}\{ [r^2-(\ell+ac)^2]\rmd r \wedge (\rmd t
-\chi\rmd \phi)\nonumber \\
    && +2rs (\ell+ac)\rmd \theta \wedge
[(r^2+a^2+\ell^2)\rmd \phi -a \rmd t]\}\ ,
\end{eqnarray}
which is the exterior derivative of the vector potential
\beq
A=-\frac{Qr}{\Sigma}(\rmd t -\chi \rmd \phi)\ .
\eeq 
Here  $\Sigma$, $\Delta$, and $\chi$ are defined by
\beq
\fl\qquad
\Sigma = r^2 +(\ell+ac)^2,\ 
\Delta = r^2-2\mathcal{M}r-\ell^2 +a^2+Q^2,\ 
\chi = a s^2 -2\ell c\ .
\eeq
Units are chosen so that $G=c=1$, so the parameters $(\mathcal{M},Q,a,\ell)$ all have the
dimension of length.
The source has mass $\mathcal{M}$, electric charge $Q$, angular momentum $J=\mathcal{M}a$
(i.e., gravitomagnetic dipole
moment) along the $\theta=0$ direction, and gravitomagnetic monopole moment
$\mu=-\ell$,
where $\ell$ is the NUT parameter. \cite{Mis,MisTau,Bon,Fei}.

The spatial 3-metric $h_{\alpha\beta}$ on the quotient space of the timelike Killing congruence associated with the Killing vector field $\xi=\partial_t$ is
\begin{equation}
\rmd l^2=h_{\alpha\beta}\rmd x^\alpha \rmd x^\beta 
=\frac{\Sigma}{\Delta}\rmd r^2 +\Sigma \rmd \theta^2 
  +\frac{\Delta s^2 \Sigma}{\Delta-a^2 s^2}\rmd \phi^2\ ,
\end{equation}
while its convenient conformal rescaling 
by the square of the lapse function 
$M^2=-\xi_\alpha \xi^\alpha = (\Delta-a^2 s^2\theta)/\Sigma$
is
\begin{eqnarray}
 \rmd \tilde l{}^2
&=& \tilde h_{\alpha\beta}\rmd x^\alpha \rmd x^\beta 
 = M^2  h_{\alpha\beta}\rmd x^\alpha \rmd x^\beta \nonumber \\
&=& \frac{\Delta-a^2\sin^2\theta}{\Delta}\rmd r^2 +(\Delta-a^2s^2) \rmd \theta^2 
   +\Delta s^2\rmd \phi^2\ .
\end{eqnarray}
Using the latter metric and its covariant derivative $\tilde\nabla$, one easily shows by direct evaluation that
both the  Ernst potential and $\Phi$
\begin{equation}
\mathcal{E}=1-\frac{2(M+il)}{r+i(\ell+ac)}\ ,\ 
\Phi= \frac{Q}{r+i(\ell+ac)}
\end{equation}
satisfy the same Scurlfree property for their gradient 
\beq
\tilde{\rm S}{\rm curl}  \, (\tilde \nabla \mathcal{E}\otimes \tilde \nabla \mathcal{E} )=0
\eeq
that occurs for the Ernst potential in the vacuum case within this family.

This is explained by the special form of the
Weyl current
\begin{eqnarray}\fl
  J^{(G)} 
&=& \frac{3\Delta s^2 a Q^2}{[r-i(\ell+ac)][r+i(\ell+ac)]^2\Sigma^{1/2}(\Delta-a^2s^2)^{3/2}}
       (\rmd r-ias \rmd\theta)\wedge \rmd\phi
\nonumber\\\fl
&=& -i \nabla \log \sigma \times  Z
\ ,
\end{eqnarray}
where 
\beq
\sigma = -\frac34 \frac{\{[r-i(\ell+ac)](\mathcal{M}+i\ell)-Q^2\}}{(\mathcal{M}+i\ell)^2}\, 
                          \frac{[r+i(\ell+ac)]}{[r-i(\ell+ac)]} 
\eeq
a factor also which appears in 
\beq
Z=M^{-2}\sigma [\nabla \mathcal{E}\otimes \nabla \mathcal{E}]^{\rm (TF)}\ ,
\eeq
and so cancels the factor inside the Scurl in the Simon tensor expression (\ref{eq:simonsigma}) when expressed in terms of $\mathcal{E}$
\beq\fl\qquad
{\rm Simon} 
= \sigma M^{-1} \Scurl (\sigma^{-1}MZ)
= \sigma M^{-1}\Scurl (M^{-1}\nabla \mathcal{E} \otimes \nabla \mathcal{E}) \ . 
\eeq
This latter Scurl expression is the same one originally used to define the Simon tensor in the vacuum case (where $\nabla\sigma \times Z=0$ and so disappears from the formula upon expansion of the first Scurl expression), and as shown in Appendix A is proportional to
$\tilde S{\rm curl}\, (\tilde\nabla \mathcal{E} \otimes \tilde\nabla \mathcal{E})$. This is forced to be zero when $z$ and $Z$ are aligned as they are here
\begin{eqnarray}
\label{eq:zZ}
z &=& 
\frac{\{[r-i(\ell+ac)](\mathcal{M}+i\ell)-Q^2\}}{[r+i(l+ac)]^2(\Delta-a^2s^2)[r-i(l+ac)]} (-\Delta \partial_r+ias \partial_\theta)
\ ,\\ 
Z &=&
\frac{3(\Delta-a^2s^2)}{\{[r-i(\ell+ac)](\mathcal{M}+i\ell)-Q^2\}} [z\otimes z]\TF
\ ,
\end{eqnarray}
In fact all the complex vectors $z$, $\nabla \mathcal{E}$, $\nabla\Phi$, and the one associated with the Killing-Yano tensor are all aligned with (i.e., proportional to) the gradient vector field
$\nabla [r+i(\ell+ac)]$. Although we have downplayed the notational distinction between contravariant and covariant tensors, by writing explicitly the index-lowered 1-form
\beq\fl\qquad
z^\flat=-\frac{\Sigma}{(\Delta-a^2s^2)}  \frac{\{[r-i(\ell+ac)](\mathcal{M}+i\ell)-Q^2\}}{[r+i(l+ac)]^2[r-i(l+ac)]} \, (\rmd r -ias \rmd \theta)\ ,
\eeq
one sees that it is proportional to this gradient (exterior derivative).

The Killing-Yano tensor \cite{Kramer} for this spacetime is naturally defined by using the principal null directions of the Riemann tensor.  The principal Newman-Penrose tetrad is
\begin{eqnarray}
l&=& \frac{r^2+a^2+l^2}{\Delta}\partial_t+\partial_r+\frac{a}{\Delta}\partial_\phi\, ,\nonumber \\
n&=& -\frac{1}{2[r^2+(l+ac)^2]}[(r^2+a^2+l^2)\partial_t+\Delta \partial_r-a\partial_\phi]\, ,\nonumber \\
m&=& \frac{1}{\sqrt{2}[(l+ac)-ir]}[\frac{\chi}{s}\partial_t-i\partial_\theta+\frac{1}{s}\partial_\phi] \ .
\end{eqnarray}
The Killing-Yano 2-form is then \cite{Kramer} (their Eq.~(35.64))
\begin{equation}
f=-(l+ac)[l\wedge n]+ir [m\wedge \bar m]\ .
\end{equation}
By taking its self-dual combination $\mathcal{K}=f+i{}^*f$, one finds 
\begin{eqnarray}\fl
\mathcal{K}
&=&
i[r+i(\ell +ac)] [\rmd t \wedge \rmd r -ias \rmd t \wedge \rmd \theta 
   +\chi \rmd r \wedge \rmd \phi -is(r^2+a^2+\ell^2)\rmd \theta \wedge \rmd \phi]\nonumber \\\fl
&=& -4i \frac{\{[r-i(\ell+ac)](\mathcal{M}+i\ell)-Q^2\}}{[r+i(\ell+ac)]^3[r-i(\ell+ac)]} 
   M \mathcal{F} 
\ ,
\end{eqnarray}
namely it is proportional to the self-dual combination (\ref{eq:Papselfdual}) of the rescaled Papapetrou 2-form (\ref{eq:Papselfdual}) associated with the timelike Killing vector.

The (symmetric)
Cotton-York tensor \cite{York} for the quotient-space geometry can be evaluated using either spatial metric since its covariant form is conformally invariant. For the rescaled metric (see \cite{cys1}), this tensor is
\beq
\tilde y= -{\tilde {\rm S}} {\rm curl} ( {}^{(3)}\tilde{\rm R}{\rm icci})
\eeq
and its only nonzero components are
\beq
\tilde y^{r\phi} = \frac{\Delta \cot \theta}{r-\mathcal{M}} \tilde y^{\theta \phi}\ ,\quad
\tilde y^{\theta \phi} =\frac{6sa^2(\ell^2+\mathcal{M}^2-Q^2)(r-\mathcal{M})}{(\Delta -a^2s^2)^5}
\eeq
The condition $\ell^2+\mathcal{M}^2-Q^2=0$ makes this tensor zero, corresponding to the conformal flatness of either spatial metric. The same is true for the static limit $a=0$, where the Cotton-York and Simon tensors coincide.

\section{Conclusions}

The gravitoelectric description of the Simon and Simon-Mars tensors which play a crucial role in characterizing the Kerr spacetime has been generalized to the class of electrovacuum spacetimes where the corresponding Kerr-Newman spacetime exhibits a similar behavior once the Simon tensor is suitably generalized to incorporate the electromagnetic field. Again the alignment of the principal null directions of the Weyl, Papapetrou and now additionally the electromagnetic field is equivalent to the vanishing of the Simon-Mars tensor, which in turn forces the generalized Simon tensor to vanish as well. The Scurl and cross-product notation and their properties also help illuminate some of the manipulations of exact solution theory whose motivation is not so obvious.

\section*{Acknowledgments}

The referees of a previous version of this article are kindly acknowledged for useful suggestions and comments.

\appendix

\section{Simon tensor for stationary vacuum spacetimes}

If $h_{ab}$ is the spatial metric, namely the Riemannian metric on the quotient 3-space of a stationary spacetime by its Killing trajectories, then a convenient conformal rescaling $\tilde h_{ab} = M^{2} h_{ab}$
by the absolute value of the norm of the timelike Killing vector field 
$M^2=-\xi_\alpha \xi^\alpha >0$ (where it remains timelike)
leads to simpler formulas for the spacetime Ricci tensor. 
Like the original definition of Simon, the formula due to Perjes \cite{Perjes85} quoted as  Eq.~(18.70) in \cite{Kramer}  relies on this conformal metric to express the Simon tensor in terms of the Ernst potential. We denote this tensor by SimonES (for the book title Exact Solutions) with a tilde (since its indices should be raised and lowered with $\tilde h_{ab}$) to distinguish it from our Simon tensor defined slightly differently using the original spatial metric
\begin{equation}
\fl\qquad
\label{simonES}
\tilde{\rm S}{\rm imonES}^b{}_a
=[\Re e\, \mathcal{E}]^{-2} \tilde\eta{}^{bcd} 
 [ \mathcal{E}_c \tilde\nabla_d \mathcal{E}_a
 -\tilde h_{a c}\tilde h^{f g}
\mathcal{E}_{[d} \tilde\nabla_{g]} \mathcal{E}_f ] \ , 
\end{equation}
where $\mathcal{E}_a=\nabla_a \mathcal{E}=\tilde\nabla_a \mathcal{E}=\tilde\mathcal{E}_a$,
$\Re e\, \mathcal{E} = -M^2$,
and all the metric and duality operations refer to the conformally rescaled 3-metric. 
(In this appendix, $\tilde\nabla$ refers to the covariant derivative associated with the metric $\tilde h_{ab}$.)
Apart from an extra factor of 2, Krish \cite{Krish} also uses this definition.

One easily sees that this tensor is symmetric, and since the second term is antisymmetric, it must exactly cancel the antisymmetric part of the first term, so only symmetric part of the first term remains and that is just the Scurl operator acting on the gradient of the Simon potential apart from the scalar multiplier, which is automatically tracefree. Thus the Simon tensor is much more efficiently represented by a formula which makes these properties obvious
\begin{eqnarray}
\fl\quad
\tilde {\rm S}{\rm imonES}_{ab}
&=&[\Re e\, \mathcal{E}]^{-2} 
   \tilde\eta^{cd}{}_{(a} \tilde\nabla_{|c|} \tilde\mathcal{E}_{b)} \tilde\mathcal{E}_d
\equiv
-[\Re e\, \mathcal{E}]^{-2}[\tilde\nabla \mathcal{E} 
    \tilde\times \tilde\nabla (\tilde\nabla \mathcal{E})]_{ab}
\nonumber \\
\fl\quad
&=& [\Re e\, \mathcal{E}]^{-2} [\tilde{\rm S}{\rm curl} (\tilde\nabla\mathcal{E} \otimes \tilde\nabla \mathcal{E})]_{ab}\ ,
\end{eqnarray}
where the last equality follows from the gradient identity (\ref{scurlXX}) in the conformal geometry.

Finally one must use the conformal transformation properties of the Scurl operator to re-express this in terms of the original spatial metric. For a symmetric covariant tensor $X_{ab} = \nabla_a \mathcal{E} \nabla_b \mathcal{E}$ which does not transform $\tilde X_{ab}= X_{ab}$, one finds
\beq
[\tilde{\rm S}{\rm curl }\, \tilde X ]_{ab}
=  [\Scurl  (M^{-1} X)]_{ab}\ ,
\eeq
so 
\beq\fl\qquad
\tilde {\rm S}{\rm imonES}_{ab}
= [\Re e\, \mathcal{E}]^{-2} [{\rm Scurl } \, 
      (M^{-1} \nabla \mathcal{E}\otimes \nabla \mathcal{E})]_{ab}
= 4 M^{-1} \,{\rm SimonMars}_{ab} \ , 
\eeq
where the last equality follows from Eq.~(\ref{simonmarsvacuum}).

Simon introduced a well-known special transformation $w=(1-\mathcal{E})/(1+\mathcal{E})$
of the Ernst potential
which has the property that it only changes the Scurl combination above by an overall factor
\beq
 \tilde{\rm S}{\rm curl}\, 
      (\tilde\nabla w \otimes \tilde\nabla w)
= \frac14 (1+w)^4\,
 \tilde{\rm S}{\rm curl}\, 
      (\tilde\nabla \mathcal{E}\otimes \tilde\nabla \mathcal{E})\ .
\eeq
As discussed in appendix B, this is actually true for any function of the Ernst potential.
The spatial dual of the original definition of Simon 
\beq
\fl
\tilde C_{bcd} 
=  4 (1-w\bar w)^{-2} (\tilde\nabla_{[d} w \tilde\nabla_{c]} \tilde\nabla_b w 
                           - \tilde h_{b[c} \tilde U_{d]})\ ,\quad
\tilde U_d= \tilde h{}^{fg} \tilde\nabla_{[d} w \tilde\nabla_{g]} \tilde\nabla_f w
\ ,
\eeq
can therefore be rewritten in the same way as the above definition
\begin{eqnarray}\fl
\frac12\tilde\eta_a{}^{cd}\tilde C_{bcd}
  &=& 2(1-w\bar w)^{-2} \tilde\eta_a{}^{cd}
           (\tilde\nabla_{[d} w \tilde\nabla_{c]} \tilde\nabla_b w
            - \tilde h_{bc} \tilde h{}^{fg} 
                   \tilde\nabla_{[d} w \tilde\nabla_{g]} \tilde\nabla_f w )
\nonumber\\ \fl
 &=& 2(1-w\bar w)^{-2} 
    [\tilde{\rm S}{\rm curl}\, (\tilde\nabla w \otimes\tilde\nabla w) ]_{ab}
\nonumber\\ \fl
&=& 2(1-w\bar w)^{-2}(1+w)^4 [\tilde{\rm S}{\rm curl}\, 
                                 (\tilde\nabla \mathcal{E} \otimes\tilde\nabla \mathcal{E})]_{ab}
\nonumber\\ \fl
 &=& 2(1-w\bar w)^{-2}(1+w)^4 [\Scurl(M^{-1}\nabla \mathcal{E} \otimes\nabla \mathcal{E})]_{ab}
\ .
\end{eqnarray}

The Scurl formula for the Simon tensor expressed in terms of the rescaled metric $\tilde h_{ab}$ has an obvious immediate consequence for the interpretation of its vanishing.
Because of the Scurl identity (\ref{scurlXXgrad}) in the conformally rescaled geometry, the differential condition
$
\tilde{\rm S}{\rm curl} (\tilde\nabla\mathcal{E} \otimes \tilde\nabla \mathcal{E}) = 0
$
is equivalent to the algebraic condition
\beq
\tilde\nabla  \mathcal{E} \tilde\times \tilde\nabla (\tilde\nabla  \mathcal{E}) = 0\ , 
\eeq
which in turn expresses the ``alignment" of the two tensors 
$\tilde \nabla (\tilde \nabla \mathcal{E})$ and $\tilde \nabla \mathcal{E}\otimes \tilde \nabla \mathcal{E}$, i.e.
\beq
[\tilde \nabla (\tilde \nabla \mathcal{E})]^{\rm (TF)} \quad \propto \quad  [\tilde \nabla \mathcal{E}\otimes \tilde \nabla \mathcal{E}]^{\rm (TF)}\ .
\eeq
This geometrical meaning has been overlooked perhaps because of the unnecessary complication in previous formulas.

For the electrovac case, the conformal transformation formula for a vector divergence
\beq
  \tilde\rmd{\rm iv}\, \tilde X = M^{-3} \div (MX)
\eeq
helps show the equivalence of Eq.~(\ref{ernst}) with the original Ernst equation (18.38) of \cite{Kramer}. Using this formula and substituting for $z$ in terms of $\mathcal{E}$ and $\Phi$ by solving Eq.~(\ref{eq:ernstpot}) in Eq.~(\ref{ernst}), one obtains
\beq
\tilde\rmd{\rm iv}\, \tilde\mathcal{E} 
  -M^{-2} \tilde h^{ab}(\tilde\mathcal{E}_a+2\bar\Phi\tilde\nabla_a\Phi)
            \tilde\mathcal{E}_b =0
\ .
\eeq

\section{Scurl and cross product identities}

In this appendix we use the simpler notation $\nabla$ for the spatial covariant derivative
and assume all fields are spatial and stationary.
A straightforward computation verifies the identity
\beq\label{scurlXX}
\Scurl (X\otimes X) 
 = \frac32\, [\SYM (X \otimes \curl X)]\TF
   -   [X\times \SYM (\nabla  X)]
\eeq
for any vector field $X$.
When the vector field is a gradient $X=\nabla f$, 
then $\curl X = \curl(\nabla f)=0$ and the covariant derivative 
$\nabla X = \nabla(\nabla f)=\SYM  (\nabla(\nabla f))$ is symmetric, so this reduces to the useful identity
(see Eq~(3.19) of \cite{cys1}) 
\beq\label{scurlXXgrad}
\Scurl(X\otimes X)  =   -   X\times(\nabla  X)
\quad {\rm if}\quad X=\nabla f\ . 
\eeq

Notice that for any function $F(f)$ of $f$, then $\nabla F(f) = F^\prime(f) \nabla f$ and one has the property that the rate of change factor passes through the Scurl operation without extra terms, expanding the Scurl by the product rule for a scalar and a tensor
\begin{eqnarray}\fl\qquad
&& \Scurl(\nabla F \otimes \nabla F)
   =\Scurl(F^{\prime 2} \nabla f \otimes \nabla f)
\nonumber\\\fl\qquad
&&=  (F^{\prime 2})^\prime \nabla f \times (\nabla f \otimes \nabla f)
   + F^{\prime 2} \Scurl(\nabla f \otimes \nabla f)
= F^{\prime 2} \Scurl(\nabla f \otimes \nabla f)
\end{eqnarray}
because of the identity $Y\times (Y\otimes Y) = 0$ valid for any vector field $Y$.

When performing actual calculations in this approach, one needs to refer to a number of Scurl and cross product identities involving vector fields and symmetric tensors where the symmetric tensors themselve arise from vector fields.
For arbitrary spatial vector fields $X$, $Y$ and $Z$ one finds by straightforward evaluation
\begin{eqnarray}
\fl\quad
\Scurl  [\SYM  (X\otimes Y)]&=& -\frac12 [X\times \SYM  (\nabla Y)+Y\times \SYM  (\nabla X)]+\nonumber \\
\fl\quad
 && +\frac34 \SYM [(X\otimes \curl Y+Y\otimes \curl X)]^{\rm (TF)}
\ ,\label{scurlXY}\\
\fl\quad
\Scurl  [\SYM  (\nabla X)]&=& - [X\times {}^{(3)}{\rm Ricci} ]
    +\frac12 \SYM [\nabla (\curl  X)]^{\rm (TF)}
\ ,\label{scurlnablaX}\\
\fl\quad
\SYM  [\nabla (X\times Y) ]^{\rm (TF)} &=& X\times \SYM  (\nabla Y) - Y\times \SYM  (\nabla X) +\nonumber \\
\fl\quad
&&+\frac12 \SYM [X\otimes \curl Y-Y \otimes {\rm curl }X ]^{\rm (TF)}
\ ,\label{scurlnablaXY}\\
\fl\quad
\SYM [X \otimes (Y\times X)]&=&Y\times[X\otimes X]
\ ,\label{XtimesYX}\\
\fl\quad
X\times \SYM [X\otimes Y]&=& -\frac12Y\times [X\otimes X]
\ ,\label{XtimesXoY}\\
\fl\quad
X \times \SYM [Y\times Z]&=&\frac12 \SYM [Y\otimes (X\times Z)+Z\otimes (X\times Y)]
\ .
\end{eqnarray}

\end{document}